\documentclass{mn2e}

\usepackage{lscape}

\title[ATCA 18\,GHz Survey] {First results from the ATCA 18\,GHz pilot survey}

\author[Ricci et al. ]{
\parbox[t]{\textwidth}{
Roberto Ricci$^{1,2}$, Elaine M.\ Sadler$^3$, Ronald D.\
Ekers$^1$, Lister Staveley--Smith$^1$, Warwick E.\ Wilson$^1$,
Michael J.\ Kesteven$^1$, Ravi Subrahmanyan$^1$, Mark A.\
Walker$^{1,3}$, Carole A.\ Jackson$^1$, Gianfranco De
Zotti$^{4,2}$ }
\vspace*{6pt} \\
$^1$Australia Telescope National Facility, CSIRO, P.O. Box 76, Epping,
    NSW 2121, Australia \\
$^2$SISSA/ISAS, Via Beirut 2--4, I-34014 Trieste, Italy \\
$^3$School of Physics, University of Sydney, NSW 2006, Australia\\
$^4$INAF, Osservatorio Astronomico di Padova, Vicolo
dell'Osservatorio 5,
I-35122 Padova, Italy \\
}

\pagerange{\pageref{firstpage}--\pageref{lastpage}}
\pubyear{2000}

\begin{document}

\maketitle

\label{firstpage}

\begin{abstract}
As a pilot study for the first all--sky radio survey at short
wavelengths, we have observed 1216\,deg$^2$ of the southern sky 
at 18\,GHz (16\,mm) using a novel wide--band (3.4\,GHz bandwidth)
analogue correlator on one baseline of the Australia Telescope
Compact Array (ATCA). We scanned a region of sky between
declination $-71^{\circ}$ and $-59^{\circ}$ with an rms noise
level of 15\,mJy. Follow--up radio imaging of candidate sources
above a 4$\sigma$ detection limit of 60\,mJy resulted in 221
confirmed detections, for which we have measured accurate
positions and flux densities. 
For extragalactic sources, the survey is roughly 70\% complete
at a flux density of 126\,mJy and 95\% complete above 300\,mJy.  
Almost half the detected sources lie
within a few degrees of the Galactic plane, but there are 123
sources with $|b|>5^\circ$ which can be assumed to be
extragalactic. The differential source counts for extragalactic
sources in the range $0.1\,\hbox{Jy} \le S_{18{\rm GHz}}\le
3\,\hbox{Jy}$ are well fitted by a relation of the form $n(S)=57\,
(S/\hbox{Jy})^{-2.2\pm0.2}\,\hbox{Jy}^{-1}\,\hbox{sr}^{-1}$, in
good agreement with the 15\,GHz counts published by Taylor et al.\
(2001) and Waldram et al.\ (2003). Over 70\% of the extragalactic
sources have a flat radio spectrum ($\alpha_{0.843}^{18}> -0.5$,
S$_{\nu}\propto\nu^{\alpha}$), and 29\% have inverted radio
spectra ($\alpha_{0.843}^{18}> 0$). The optical identification
rate is high: 51\% of the extragalactic sources are identified
with stellar objects (candidate QSOs), 22\% with galaxies and only 27\%
with faint optical objects or blank fields.
\end{abstract}



\section{Introduction}
The radio--source population above 5\,GHz has not yet been well
studied.  This is mainly because large radio telescopes typically
have fields of view of a few arcmin at high frequencies, making it
extremely time--consuming to carry out large--area surveys.
Measuring the high--frequency properties of extragalactic radio
sources is also crucial for interpreting the high--sensitivity and
high--resolution maps of the Cosmic Microwave Background (CMB)
provided by WMAP (Bennett et al.\ 2003) and forthcoming missions
like ESA's Planck satellite (Tauber 2001).
Current analyses (De Zotti et al. 1999a) 
indicate that the best regions for
high--resolution CMB mapping lie near 100\,GHz, and that the main
contaminant is likely to be the fluctuations due to individual
foreground radio sources.

The brightest high--frequency radio sources are expected to be
blazars (high and low-polarization flat-spectrum quasars and BL
Lac objects) and galaxies 
with highly--inverted radio spectra such
as the GHz Peaked Spectrum (GPS) sources (O'Dea 1998). The
Michigan 8\,GHz ($\lambda$3.8\,cm) survey covered
$\sim2600$\,deg$^2$ of sky and detected 55 sources down to a
limiting flux density of 0.6\,Jy.  These sources were found to be
different from those detected in low--frequency surveys, with 70\%
having flat or inverted spectra.  More than 80\% had optical
counterparts on the Palomar Sky Survey, most of them QSOs (Brandie
\& Bridle 1974; Brandie, Bridle \& Kesteven 1974).

The first blind radio survey above 8\,GHz was carried out by
Taylor et al.\ (2001) with the Ryle Telescope. They observed
63\,deg$^2$ of sky and detected 66 sources to a limiting flux
density of 20\,mJy at 15.2\,GHz. A significant fraction (20\%) of
the sources they detected would not have been predicted by a
simple extrapolation from lower--frequency surveys. Two--thirds of
their sources had flat or inverted spectra. 
The source density observed by Taylor et al.\ was also a factor of
1.3--2 lower than predicted by theoretical models based on 
low-frequency radio luminosity functions (Dunlop \& Peacock 1990;
Toffolatti et al. 1998). 

The 15\,GHz survey was extended by Waldram et al.\ (2003) 
 to 520\,deg$^2$ in three regions of sky targeted by the
Very Small Array (Watson et al.\ 2003). They found 465 sources above a
completeness limit of $\sim$25 mJy in their main survey,
although a total of $\sim$760 sources were detected down to a limiting
flux density of 10 mJy. 



Preliminary radio source counts at 30\,GHz have been
measured by the DASI (Leitch et al.\ 2002), VSA (Grainge et al.\ 2003)
and CBI (Mason et al.\ 2003) groups, and a catalogue of 208 point
source candidates stronger than about 1\,Jy from the first year of
WMAP data (five frequency bands from 23--94\,GHz) has recently
been published by Bennett et al.\ (2003).

In this paper, we present the results of a 18\,GHz Pilot Survey
at the Australia Telescope Compact Array 
(ATCA)\footnote{See www.narrabri.atnf.csiro.au and 
Journal of Electrical and Electronics Engineering, Australia,
Special Issue, Vol. 12, No. 2, June 1992}. 
The pilot survey covered 1216\,deg$^2$ of sky 
in the region between $\delta = -71^{\circ}$ and $-59^{\circ}$ 
to a (4$\sigma$) limiting flux density of 60\,mJy.
In the follow-up observations we confirmed 221 sources. 

In Section 2, we briefly describe the survey strategy and
give some details of the observing run. Section 3 describes the
data--reduction strategy for the scan data, which has some novel
features. Follow-up synthesis imaging observations of the detected
sources are described in Section 4, and their data reduction in
Section 5.  Section 6 presents the follow-up results, Section 7
a comparison with surveys at other frequencies, Section 8 the
optical identifications and Section 9 summarizes our conclusions.


\section{Survey strategy and observations}

An analogue wide--band prototype correlator
originally developed for the Taiwanese CMB instrument AMiBA
(Lo et al. 2001) was linked to two ATCA
antennas to produce a two--element interferometer with 3.4 GHz
bandwidth.
The two 22m antennas were placed on the shortest ATCA baseline (30\,m),
giving a fringe spacing of $\sim90$\,arcsec at the centreband
frequency of 18\,GHz.
The correlator produced 16 lag channels corresponding  (after the Fourier
inversion) to 8 frequency channels, each with a bandwidth of 425\,MHz.
This large total bandwidth allowed us to obtain 
high signal-to-noise measurements in the very short integration time used
(0.2 sec/beam).
Such rapid scanning was a design feature of the ATCA implemented to perform 
mosaic observation over large areas. This combination of wide 
bandwidth and fast scanning has made it possible to observe a large fraction
of the sky despite the small ATCA primary beam
at 18\,GHz ($\hbox{HPBW}=2.3$\,arcmin).
Only a single linear polarization was implemented in this 
correlator configuration.


No delay tracking had been built for this analogue correlator 
(AMiBA antennas are on a common tracking mount) so we designed an 
observing mode in which all observations are taken 
on the meridian where the delay for an east--west 
interferometer is zero.

The scanning strategy was as follows:
the two ATCA antennas point in the same direction and scan the
sky synchronously up and down between $\delta = -70^{\circ}$ and $-60^{\circ}$
in elevation along the meridian with a scan rate
of 10$^{\circ}$ per minute (the maximum speed the antennas
can slew in a closed loop in elevation).
The Earth's rotation then allows us to cover a strip of
the sky in raster scan mode.
Antenna coordinates were sampled at 10\,Hz. With an 
integration time of 80 ms, the antenna pointing was recorded
every 0.8\,arcmin.

The pilot survey run took place on 2002 September 13--17 during one
observing period of 96 hours. 

\section{Survey data reduction}

The antenna positions (obtained from the telescope encoders) and
fringe visibilities (produced by the analogue correlator) were
stored in data files. Custom MIRIAD (Sault, Teuben \& Wright 1995) 
tasks were written to flag out bad data
(e.g.\ antenna pointings more than 30\arcsec\ from one another)
and to calibrate the visibilities in absolute flux and bandpass.

The next stage involved searching for peaks in the fringe
visibility amplitudes above a given signal-to-noise ratio in the
time-ordered raster scans, to produce a candidate source list.
This was done using the following steps:

\begin{enumerate}
\item
We calculated the weighted complex mean of the 8 separate frequency
channels to form a single 3.4 GHz wide band.  
The bandwidth smearing (Thompson 1999) in the worst case (a source 
at half power point of the primary beam) is 1/6 of the fringe spacing
and can be ignored. Expressed in a different way our 15\% bandwidth
corresponds to a 5.6m variation in the 30m spacing used, which is a 
small fraction of the 22m dish diameter.  
\item
We apply a beam shape Gaussian ($\hbox{FWHM}=2^{\prime}$.3) filter
acting as a weighted running mean for this short baseline 
to smooth out fluctuations (generated by receiver noise)
on spatial scales less than the antenna primary beam.
\item
In the plane defined by the real an imaginary parts of the 
fringe visibility we determined the dispersion along each axis.
Points lying at $> n \sqrt{\sigma_{x}^2 + \sigma_y^2}$ are selected 
as candidate sources at $n \sigma$ level. In general a source generates 
up to 4 contiguous points along the time series corresponding to 
the sampling by different parts of the beam in steps of 0.8\,arcmin.
The candidate source position corresponds to the maximum visibility 
amplitude in this sequence.    
\item
We created a list of candidate point--source detections
at a range of signal-to-noise ratios (3, 3.5, 4 and 5\,$\sigma$)
to explore the optimum detection strategy. 
\end{enumerate}

The raw candidate source lists were then merged into a single list.  
The number of $4\sigma$ candidate sources for the entire survey
was 574. The measured rms noise in the scans was $\sigma_{\rm rms}=15\,$mJy.

We performed a Monte Carlo simulation to estimate
the number of spurious peaks generated by Gaussian noise at a given
signal--to--noise ratio (see Table~\ref{tab:spur}).
The same number of real and imaginary parts of the fringe
visibilities for each frequency channel was generated 30
times by sampling a Gaussian distribution with the same mean and
standard deviation as the actual distributions.
A cutoff value of ${\rm S/N}= 4$ was chosen as a good trade-off,
since lower values yield too many spurious peaks while for
${\rm S/N}= 5$ we reject too many sources which are genuine
detections. Since the S/N in the follow-up is very high compared with
the 0.2 sec survey integration, we could afford to explore the 
detection statistics by going well into the region of spurious 
detections due to noise.  

\begin{table}
\begin{center}
\begin{tabular}{ccc}
\hline
S/N   &    $\hbox{N}_{\rm exp}$   &  $\hbox{N}_{\rm sim}$  \\
\hline
3     &    542         &  529$\pm$21  \\
3.5   &    111         &  104$\pm$10  \\
4     &     24         &   16$\pm$4   \\
5     &     5          &    0.13$\pm$0.35 \\
\hline
\end{tabular}
\caption{Comparison between the actual number of peaks in a raster scan
during one test run ($\hbox{N}_{\rm exp}$), and the number 
of spurious peaks predicted
by a Monte Carlo simulation of the same scan ($\hbox{N}_{\rm sim}$) as
described in the text.}
\label{tab:spur}
\end{center}
\end{table}

\section{Follow-up observations}

Follow-up synthesis--imaging observations of the ${\rm S/N}\geq4$
candidate sources were carried out on 2002 October 8--12 with the
ATCA in its hybrid H168B configuration, which has antennas
along the new 214\,m Northern Spur as well as the existing east--west
railway track.
At the time of this run, only three antennas
had been equipped with low--noise 12\,mm receivers, so only three baselines
were available for imaging.  We therefore observed each object over
a wide range in hour angle to maximize $uv$ coverage.
The angular resolution depends on the source declination. The median
major and minor axis values are 30 and 13\,arcsec respectively. 
The average rms noise in synthetized maps was $\sigma\simeq 1.5$ mJy.   

We used the standard ATCA correlator, allowing us
to observe in two spectral windows centred at 17.2 and 18.7\,GHz
each with a bandwidth of 128 MHz.
At 18\,GHz, the standard ATCA primary calibrator $1934-638$
is too faint to provide a good absolute flux density scale, and
we used the planet Mars as our primary calibrator.  
The fringe--visibility amplitude dependence
on spatial frequency has been accurately modelled for the ATCA.
The Mars model we used is described in Ulich (1981) and Rudy (1987).
As in Ulich (1981), the seasonal variation of Mars brightness 
temperature is also taken in account by the model. 

\section{Follow-up data reduction}

\subsection{Follow--up 18\,GHz images}
Reduction of the follow--up 18\,GHz images was done in the standard
way using the MIRIAD software package (Sault, Teuben \& Wright 1995).
The target sources and calibrators were imaged in total intensity (Stokes I)
and circular polarization (Stokes V).  The sources were assumed to have zero
circular polarization, and the Stokes V maps were used to estimate the rms
noise.  We used the MIRIAD Multi--Frequency Synthesis (MFS) algorithm at the
Fourier inversion stage, since this improves $(u,v)$ coverage by
simultaneously Fourier--transforming the visibility data from both
observing frequencies.  After the Fourier inversion the maps
were CLEANed, and the resulting clean components
were restored using an elliptical Gaussian beam.

We used the maximum pixel value in the restored Stokes I map together
with the V Stokes rms noise 
to calculate the map signal-to-noise (hereafter S/N) ratios.
A source was considered to be detected when its S/N value was above 5.

The number of sources above this detection threshold was 282 out of
574 candidates. However, many candidate objects ($\sim 20\%$) were
multiple detections of the same source, so that the actual number of spurious
sources in the survey was smaller. These multiple observations were used to 
estimate data quality.

\subsection{Flux density and position measurements}

We reimaged the detected sources in all four Stokes parameters,
applying the primary beam correction to restored maps and a natural
weighting to the fringe visibilities before the Fourier inversion.
The primary beam correction was only applied at this last stage,
because the increase in noise near the map border complicates the detection
statistics.

All but five of the confirmed (S/N$>5$) sources were well--fitted by a
single elliptical Gaussian. 

We created a final source list by removing all sources which had the
same position (within the errors) as a previously--listed one.
This decreased the final number of detected sources from 282 to 221,
and most of the subsequent analysis was done on this list of objects.
Source parameters determined by the MIRIAD
IMFIT algorithm\footnote{Described in wwwatnf.csiro.au/computing/software/miriad/ taskindex.html} for the confirmed sources are presented in Table 2.
Any source having an angular size larger than 10 arcsec was dubbed (R) 
meaning ``significantly resolved''. Because of angular size measurement
errors, point-like sources at the resolution of follow-up images could 
appear resolved.   

\subsection{Polarization}
Determining the level of polarization in foreground radio sources
is critically important for the study of polarization anisotropies
in the cosmic microwave background (e.g. Hu, Hedman \& Zaldarriaga
2003). De Zotti et al.\ (1999b) and Mesa et al. (2002) estimate
that extragalactic sources will not be a strong limiting factor in
measurements of E-mode CMB polarization, but there are few existing
measurements of the polarization of radio galaxies and quasars at
frequencies above 5\,GHz (Ricci et al. 2004).

For all the sources and calibrators we produced 
polarized intensity {\it p}, fractional polarization
{\it m} and position angle {\it PA} maps and the relative error maps 
after clipping (i.e. blanking) pixels with S/N ratios below 3. 
The peak flux in the Q, U and I maps was
then measured at the I map peak position using the MIRIAD MAXFIT task 
(see the MIRIAD web page in the previous footnote), 
and the fractional linear polarization estimated through the
relation:
\begin{equation}
m=\frac{\sqrt{Q^2+U^2}}{I}
\end{equation}

\setcounter{table}{2}
\begin{table}
\begin{center}
\begin{tabular}{lrrrl}
\hline
Source       &  \multicolumn{1}{c}{$b$} & $S_{\rm peak}$  & Frac.  & ID \\
             & (deg)&       (mJy)     & pol. &    \\
\hline
1147-6753   &  $-$5.9 &   1150  &    2.0\%  & QSO \\ 
1236-6150   &  $+$1.0 &    422  &    0.7\%  & Galactic  \\
1243-6254   &  $-$0.06&    787  &    2.8\%  & Galactic  \\
1310-6245   &  $+$0.03&    379  &    5.0\%  & Galactic? \\
1311-6246   &  $-$0.01&   1264  &    0.5\%  & Galactic? \\
1337-6508   &  $-$2.7 &    765  &    5.1\%  & Extragalactic \\
1346-6021   &  $+$1.8 &   1503  &    7.9\%  & Radio galaxy \\ 
1400-6209   &  $-$0.4 &    231  &   15.1\%  & Galactic  \\
1647-6438   &  $-$12.5&    534  &    6.9\%  & Radio galaxy \\
1803-6507   &  $-$19.6&    945  &    2.0\%  & Radio galaxy \\
1824-6717   &  $-$22.4&    105  &   16.2\%  & Radio galaxy? \\
2303-6806   &  $-$46.0&    848  &    3.7\%  & QSO \\
2312-6607   &  $-$48.1&    102  &    2.0\%  & QSO? \\
\hline
\end{tabular}
\caption{Linear polarization detections at 3 $\sigma$.} 
\vspace*{-0.5cm}
\label{tab:pol}
\end{center}
\end{table}

Only 13 of the 221 follow-up confirmed sources were found to have
a linear polarization intensity $p=\sqrt{Q^2+U^2}$ above
3$\sigma_p$, where $\sigma_{p}$ is typically 6\,mJy (Table~\ref{tab:pol}). 
Six of these have $|b|>5^\circ$ and appear to be extragalactic.
Two extra-galactic objects in the $|b|<5^{\circ}$ Galactic strip were also 
detected in polarization.    

\begin{figure}
\begin{center}
\includegraphics{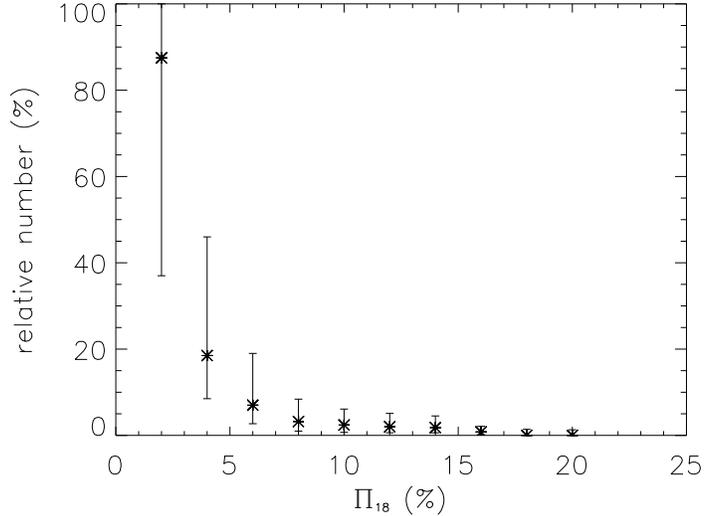} \vspace{7cm}
\caption{Fractional polarization distribution at 18 GHz given by the 
ratio between the number of polarization detected sources and the number 
of sources that could be detected at this level. 
Errors bars provide the maximum range permitted by Poissonian errors
in source sample size per bin.}
\label{fig:cumpol}
\end{center}
\end{figure}

In Fig.~\ref{fig:cumpol} we show the fraction of 18 GHz polarized sources
that were actually detected below a certain polarization level inside 
the sample of sources that could be detected below that polarization
level. This statistical method, similar to the one used by Auriemma et. al
(1977) to study radio luminosity functions, gives an estimate of 
the fractional polarization distribution. Error bars indicate the 
maximum range inside each polarization sub-sample.     
The first polarization level is strongly affected by small number 
statistical Poissonian errors, having 7 polarization detections out of
8 sources in which polarization could be detected.
If we take only the extragalactic sources into account we have a 
median polarization of $\sim2.1\%$. The average flux density of all 
extragalactic sources contributing to the median is 1.0 Jy.

\begin{figure}
\begin{center}
\includegraphics{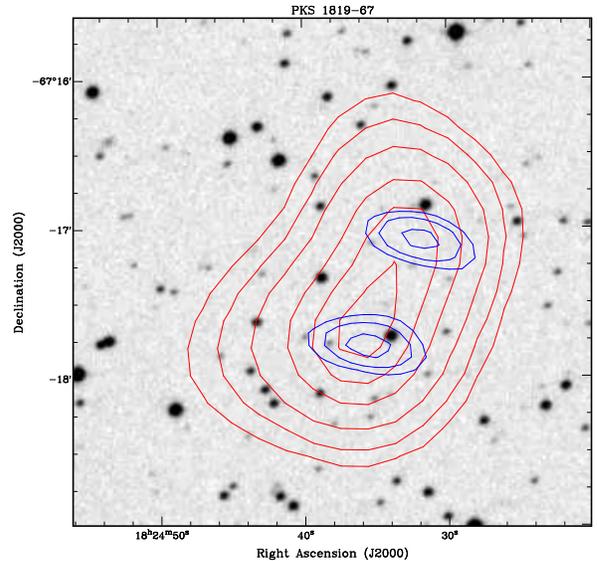} \vspace{7.2cm}
\caption{Radio contours for PKS\,1819$-$67 at 843\,MHz (larger scale) and
18\,GHz (smaller scale) overlaid on a blue SuperCOSMOS image. This source has
the highest linear polarization (16.2\%) measured in our pilot study.   }
\vspace*{-0.8cm}
\label{fig:pks1819}
\end{center}
\end{figure}

We also draw attention to the extragalactic source 1824$-$6717
(=PKS\,1819$-$67), which has the highest fractional polarization
so far measured in our sample at 16.2\%. 
Since PKS\,1819$-$67 is a widely--spaced (47\,arcsec)
double, with extended radio emission seen at lower frequencies
(Fig.~\ref{fig:pks1819}), it is unlikely to be a distant
gravitationally--lensed radio galaxy. 
We are probably seeing a pair of radio hotspots in a powerful FRII
radio galaxy. Since such sources are identified with intrinsically 
bright elliptical galaxies and  
no optical counterpart is seen on the DSS,
the galaxy redshift should be $z>1$.

\section{Distribution of sources}

\subsection{Distribution of the detected sources on the sky}

\begin{figure}
\begin{center}
\includegraphics{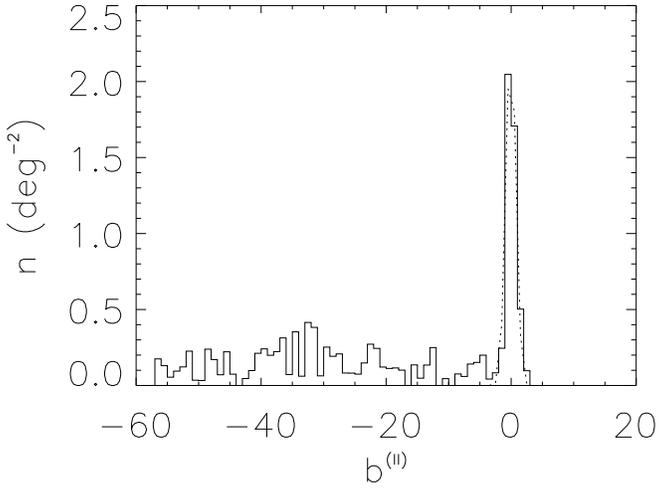} \vspace{6.5cm}
\caption{Histogram of source density in 1-degree bins
of Galactic latitude $b$ (corrected for the area surveyed). The peak near
$b$=0$^\circ$ corresponds to Galactic disk sources
such as HII regions. The surveyed area (see Fig.~\ref{fig:galcoord})
only extends to $b=+3^{\circ}$ so the apparent cut-off in sources at 
$b>+3^{\circ}$ is not real.}
\label{fig:galden}
\end{center}
\end{figure}

\begin{figure}
\begin{center}
\includegraphics{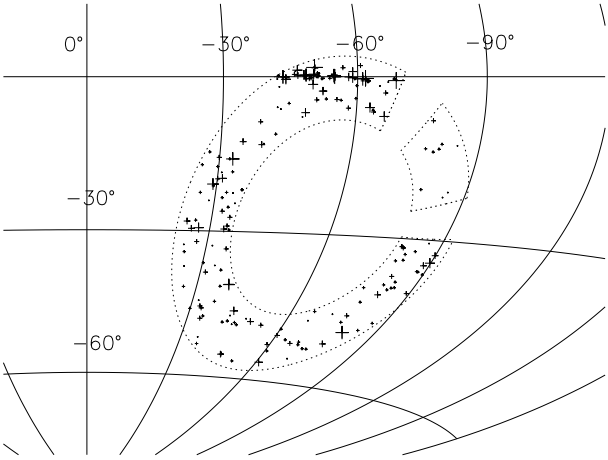} \vspace{7cm}
\caption{Equal area Aitoff projection sky map in galactic
coordinates of the confirmed 18\,GHz sources. Larger crosses
correspond to brighter sources.}
\label{fig:galcoord}
\end{center}
\end{figure}

As can be seen from Figs.~\ref{fig:galden} and \ref{fig:galcoord},
many of the confirmed 18\,GHz sources lie in or near the Galactic
plane. 

Using the source density histogram in Fig.~\ref{fig:galden}, 
we estimated the number of galactic objects that are expected to 
be found in the extragalactic survey region. The galactic plane shape in 
the source density histogram was fitted with a single Gaussian profile.
The best fit parameters resulted to be $n_{\rm peak}$=2.3 deg$^{-2}$,
$b_0$=0 and $\sigma$=0$^{\circ}$.8, where $n_{\rm peak}$
is the source density peak, $b_0$ is the galactic latitude value at the 
source density peak and $\sigma$ is the Gaussian profile width. 

In the discussion which follows we assume that the 123 sources
which have $|b|>5^{\circ}$, and which lie more than 5.5\,degrees from
the LMC centre ($\alpha=5^{\rm h}\, 23^{\rm m}\,
34^{\rm s}.7$, $\delta = -69^\circ\, 45'\, 22''$, J2000;
$l=280.47^\circ$, $b=-32.89^\circ$) are extragalactic.
When split in this way, the final sample of confirmed sources in
Table 3 contains 84 Galactic, 14 LMC and 123 extragalactic
sources. 

         
{\rm \subsection{Survey sensitivity}

Because this pilot survey did not achieve uniform coverage we made 
a quantitative assessment of the survey sky coverage through
a survey sensitivity map, i.e. a sky map representing
the limiting flux density value at every point of the sky covered
by the survey. To create this map, the survey sky strip
($-71^{\circ} < \delta < -59^{\circ}$) was gridded into
30\,arcsec$\times$30\,arcsec pixels, and the $\sim2736890$
survey sample pointings were projected onto this
1800$\times$24000-pixel grid.
To account for the antenna primary-beam pattern,
this image was convolved with a 7$\times$7-pixel
discretized Gaussian function ($\hbox{HPBW}=2.3\,{\rm arcmin}$).
The convolved pointing number per pixel was then multiplied by
the time $t_{\rm cross}$ taken by a single beam to sweep across a pixel
($t_{\rm cross}=0.28\,$sec) to get the pixel observing
time. The antenna system temperature ($T_{\rm sys}=78\,$K), the
antenna gain at 12mm ($G=0.10\,\hbox{K}\,\hbox{Jy}^{-1}$), the
observing time per pixel $\tau$ and the correlator band-width
($\Delta\nu= 3.4\,$GHz) were used to compute the pixel rms
sensitivity $\Delta S_{\rm rms}$.
A quantitative representation of the
sensitivity map is offered by the effective area $A_{\rm eff}$ as
a function of the limiting flux density (Fig.~\ref{fig:Aeff}),
which is simply given by the cumulative number of pixels for which
$4\times \Delta S_{\rm rms}\le S_{\rm lim}$ multiplied by the
single pixel area.

\begin{figure}
\begin{center}
\includegraphics{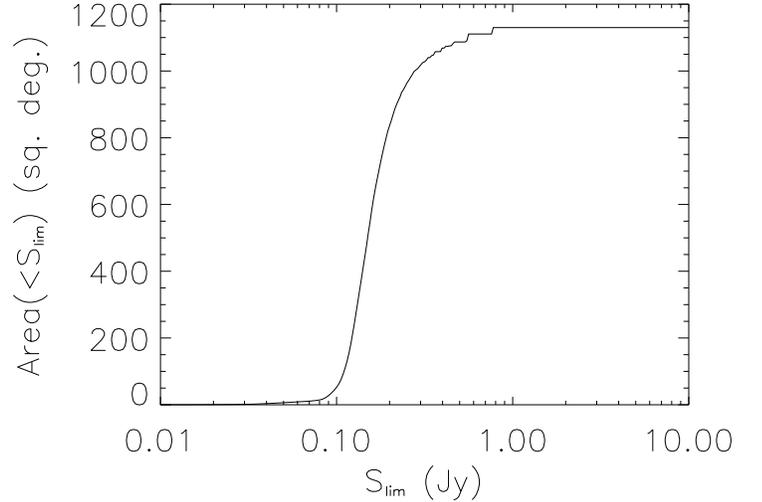} \vspace{7cm}
\caption{Effective area A$_{\rm eff}$ surveyed in the extragalactic region
of the pilot survey scans as a function of the limiting flux density
($S_{\rm lim}=4\times \Delta S_{rms}$).  
The maximum surveyed area outside the Galactic latitude range ($|b^{\rm II}| > 
5^{\circ}$), excluding the LMC region, is A$_{\rm tot}=1130\,\hbox{deg}^2$.}
\label{fig:Aeff}
\end{center}
\end{figure}

\subsection{Differential counts of extragalactic sources}
Here, we determine source counts for the extragalactic population
at 18\,GHz.  We exclude Galactic sources, since many have large
angular size and their integrated flux densities are underestimated
in the follow--up snapshot images.

The counts were logarithmically binned in flux density starting
from S$_{\rm lim}$=100\,mJy with a common binsize of 0.2 dex. A
limiting flux density of 100\,mJy was chosen because  
only a handful of objects were detected below this flux density.

As the effective surveyed area $A_{\rm eff}$ drops with decreasing
limiting flux density $S_{\rm lim}$ below $S_{\rm lim} \simeq
1\,$Jy, the surface density of sources in the (small) flux density
interval $\Delta S = S_2 - S_1$, $n(S)\Delta S$, is given by:
\begin{equation}
n(S)\Delta S = N_{\rm obs}(>S_1) \frac{1}{A_{\rm
eff}(>S_{1})}-N_{\rm obs}(>S_2)\frac{1}{A_{\rm
eff}(>S_{2})} \ , \label{countscorr}
\end{equation}
where $N_{\rm obs}(>S)$ is the number of counted sources brighter
than $S$. This procedure is equivalent to the more commonly used 
method of calculating the effective area over which each source
could have been observed and summing the inverse area in the 
flux density bin of interest (Katgert et al. 1973).  

The computed source counts need to be corrected for incompleteness.
The fraction of sources with true flux $S > 4\sigma$,
that should be in our sample but are missed because their observed flux
$S_{\rm obs} < 4\sigma$ is given by:

\begin{equation}
F_{\rm missed}={\displaystyle\int_{\sigma_{\rm min}}^{S/4} d\sigma \ A(\sigma)\ 
\displaystyle\int_{0}^{4\sigma} dS_{\rm obs}\ {e^{-\frac{1}{2}\left(\frac{S-S_{\rm obs}}{\sigma}  \right)^2} / {\sqrt{2\pi}\sigma}}\over {\displaystyle\int_{\sigma_{\rm min}}^{S/4} d\sigma \ A(\sigma)}}
 \ . \label{missed}
\end{equation}
The values of $F_{\rm missed}$ as a function of $S$ are shown in 
Table~\ref{tab:counts} together with their negative and positive
errors.

\setcounter{table}{4}
\begin{table}
\begin{center}
\caption{Differential counts of extragalactic sources, corrected for 
incompleteness.}
\begin{tabular}{rrrrr}
\multicolumn{1}{c}{$S$ (Jy)} &  \multicolumn{1}{c}{$F_{\rm missed}$ (\%)} & \multicolumn{1}{c}{$n(S)$ (Jy$^{-1}$ sr$^{-1}$)} 
& \multicolumn{1}{c}{neg. err.} & \multicolumn{1}{c}{pos. err.}  \\
\hline   
      0.126 &    31.4 &   9298   & 5735   & 33452    \\ 
      0.200 &    14.3 &   1175   &  286   &   580    \\
      0.316 &     4.4 &    412   &   97   &   125    \\
      0.501 &     2.0 &    174   &   48   &    63    \\
      0.794 &     3.3 &     57   &   21   &    31    \\
      1.259 &     0.0 &     20   &    9.6 &    16    \\
      1.995 &     0.0 &      9.5 &    5.2 &     9.2  \\ 
      3.162 &     0.0 &      2.0 &    1.6 &     4.6  \\ 
\hline
\end{tabular}
\label{tab:counts}
\end{center}
\end{table}  

 
 
We have used the lowest possible angular resolution in the survey 
to minimize the selection against extended sources, but any sources
extended more than a few minutes of arc could be missed at the survey 
limit. This will have a very small effect on the estimate of the total 
source counts.    
 
The uncertainties on differential counts were estimated by summing
in quadrature the Poisson errors (Gehrels 1986) and the
uncertainties in the effective area. The results are plotted in
Fig.~\ref{fig:dsc}, where we also show the least square linear fit
given by:
\begin{equation}
n(S)=57\left(\frac{S}{\mbox{Jy}}\right)^{-2.2\pm0.2}
\mbox{Jy}^{-1} \mbox{sr}^{-1}
\end{equation}
Our counts are in good agreement with those by Waldram et al.\ (2003)
(9C survey) at 15\,GHz (Fig.~\ref{fig:ndsc}), as well as
with the preliminary VSA (Grainge et al. 2003),
DASI (Leitch et al. 2002),
and WMAP counts (Bennett et al. 2003) if we assume a radio spectral index
$\alpha=0$.

\begin{figure}
\begin{center}
\includegraphics{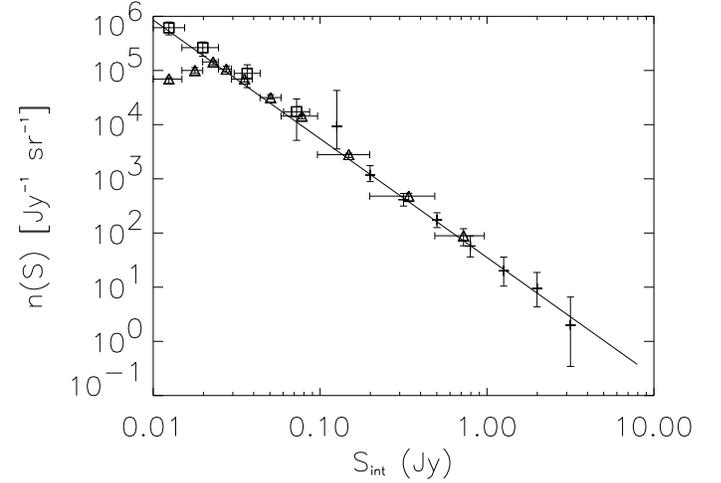} \vspace{7cm}
\caption{Effective area--corrected and completeness-corrected
differential source counts of
the extragalactic population at 18 GHz are presented together with
9C main survey (triangles) and deeper survey (squares) counts by
Waldram et al. (2003). The solid line shows the least-squares
linear fit to our counts. The two lowest flux density
 bins (triangles) of the 9C main survey are underestimated due to 
incompleteness.}
\label{fig:dsc}
\end{center}
\end{figure}

\begin{figure}
\begin{center}
\includegraphics{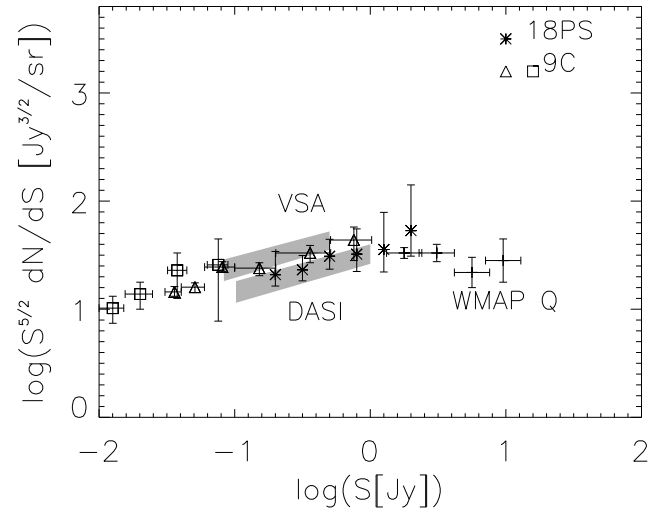} \vspace{7cm}
\caption{Euclidean normalized differential source counts from
our pilot survey (18PS) compared with VSA, DASI, WMAP band Q (41 GHz) and
9C main (triangles) and deeper (squares) surveys. Modified version
of Fig. 13 from Bennett et al. (2003).}
\label{fig:ndsc}
\end{center}
\end{figure}

\section{Comparison with data at lower radio frequencies}

\subsection{Position comparison}
Twenty of the objects detected in the 18\,GHz pilot survey are
strong compact radio sources which have previously been observed
by Ma et al.\ (1998) with Mark\,III VLBI at 2.3 and 8.4\,GHz as
part of a program to tie down the International Celestial Reference
Frame. 
The mean offset is $+$0.5\,arcsec
in RA and $-$0.2\,arcsec in Dec, with standard deviations of 3.0
and 1.4 arcsec respectively. These standard deviations are
larger than the formal uncertainties (2.0\,arcsec in RA and
1.1\,arcsec in Dec) measured in the Miriad source--fitting, and
reflect the true external errors in our position measurements.

We also cross-matched our extragalactic sub-sample with the 0.843\,GHz
SUMSS (Mauch et al.\ 2003) and 4.85\,GHz PMN (Gregory et al.\ 1994)
catalogues.  All the 123 18\,GHz extragalactic sources
have a PMN counterpart, and all 102 sources lying in the region of overlap
with SUMSS have a SUMSS counterpart. 
    
\subsection{Spectral index distribution}

Fig.~\ref{fig:ffsc} shows that the low-frequency (S$_{0.843}$) 
and high-frequency (S$_{18}$) flux densities are uncorrelated.
It is important to stress this point because in many CMB experiments
the extra-galactic radio source foreground is computed starting from 
low-frequency source catalogues and assuming a distribution of spectral 
indices around a mean value to model the source population at high 
frequency ($>$30 GHz), but in doing this, a correlation between low- 
and high-frequency flux density is implicitly assumed.      

\begin{figure}
\begin{center}
\includegraphics{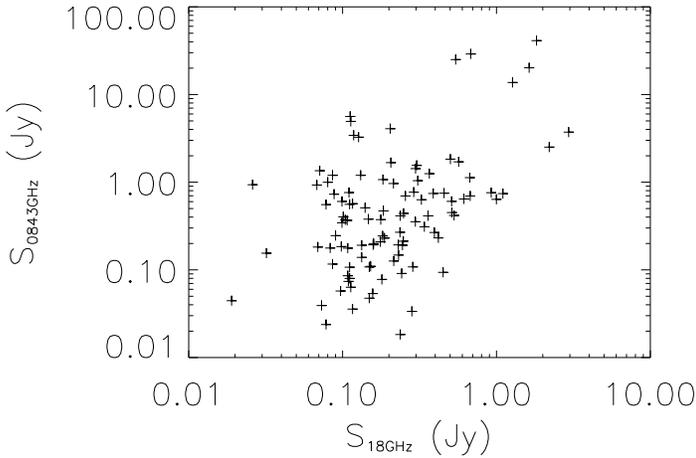} \vspace{7.5cm}
\caption{S$_{0.843}$ vs. S$_{18}$ scatter plot. No evidence of a significant
correlation between low- and high-frequency fluxes is present.}  
\vspace*{-0.5cm}
\label{fig:ffsc}
\end{center}
\end{figure}
 
Fig.~\ref{fig:spixhisto} shows the spectral index
distribution ($\alpha$ where $S_{\nu}\propto \nu^{\alpha}$) for
the extragalactic population. Most spectral indices are in
the range $-1.5 \le \alpha \le 0.5$. Over 70 \% of the extragalactic 
sources have a flat radio spectrum ($\alpha_{0.843}^{18} > -0.5$) and 
29\% have an inverted spectral index ($\alpha_{0.843}^{18} > 0$).

\begin{figure}
\begin{center}
\includegraphics{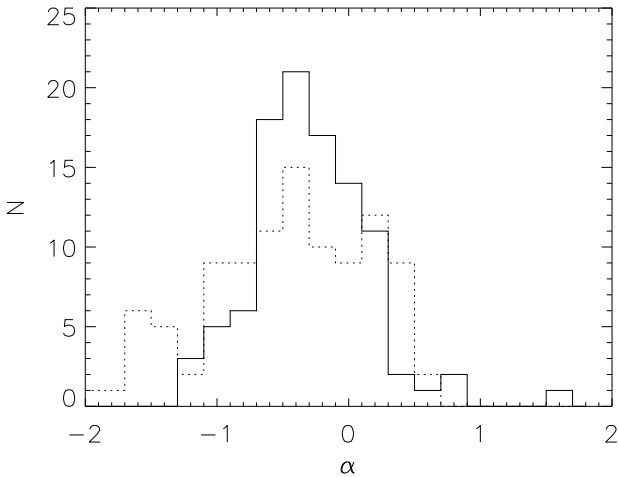} \vspace{7.5cm}
\caption{Spectral index distributions of the extragalactic sample.
The solid line shows a histogram of
the spectral indices $\alpha_{0.843}^{5}$ calculated from
SUMSS at 843\,MHz (median value=$-$0.22), and the dashed
line $\alpha_{5}^{18}$ calculated from PMN at 5\,GHz 
(median value=$-$0.29).}
\vspace*{-0.5cm}
\label{fig:spixhisto}
\end{center}
\end{figure}

The $\alpha_{0.843}^{18}$ spectral index distribution we obtained is in 
good agreement with the Michigan 8 GHz survey (70\% flat or inverted sources)
and Taylor et al. (2001)'s distributions (2/3 of the sources flat or 
inverted). However, the 9C (Waldram et al. 2003) spectral index 
distribution between 1.4 GHz NVSS and 15 GHz flux densities, 
which goes about 4 times fainter, contains 
almost equal numbers of flat and steep spectrum sources (55\% flat or 
inverted), suggesting that the ratio of flat to steep spectrum sources
decreases at lower flux density.
The same behaviour is found for the spectral index 
statistics of a lower frequency survey such as the 5 GHz selected 
1 Jy sample (54\% of flat or inverted spectrum sources, Stickel et al. 1994). 
Theoretical predictions based on low frequency selected 
radio luminosity functions (e.g. Dunlop \& Peacock
1990) and simple spectral scaling laws (flat spectrum $\alpha = 0$ and 
steep spectrum $\alpha = -0.7 $) tend to overestimate the source 
density at high frequency and fail to reproduce the observed 
proportionality between flat and steep spectrum populations.
High-frequency turn-over in radio source spectra due to synchrotron 
electron ageing needs to be modelled as we are at reasonably high
observing frequency and some sources lie at high redshift.   

To better characterise the spectral behaviour of the extra-galactic
population we also present an $\alpha_{5}^{18}$ vs. $\alpha_{0.8}^{5}$
scatter plot in Fig.~\ref{fig:spixsc} for the 101 extragalactic 
objects having a flux density measurement at 0.8, 5 and 18 GHz.

We found that 15 sources classified as inverted at low frequency 
($0 < \alpha_{0.8}^{5} < 0.5$) are also inverted at high 
frequency ($\alpha_5^{18} >0$).
No sources out of the 6 classified as High Frequency Peaker candidates 
having $\alpha_{0.8}^{5} > 0.5$ show a rising spectrum between 5 and 18 GHz.
14 sources being flat between 0.8 and 5 GHz ($-0.5 < \alpha_{0.8}^{5}
< 0$) become inverted at higher frequency.        
Most of the steep spectrum ($\alpha_{0.8}^{5} < -0.5$) sources (except 2
objects) stay steep even in the high frequency range. 

In the $\alpha_{5}^{18}$ histogram in Fig.~\ref{fig:spixhisto} there are 
9 sources in the 0.4--0.6 bin but only 2 are present in the same spectral 
range in the $\alpha_{0.8}^{5}$ histogram. The radio spectra of these 
objects are turning up at higher frequency. It is worth noticing that 
the PMN flux densities could be overestimated as the PMN survey 
beam was so large (3 arcmin) that it is possible that more than one source 
is included inside the beam area. This would flatten the 0.843--5 GHz 
spectral indices and steepen the 5--18 GHz ones.            

\begin{figure}
\begin{center}
\includegraphics{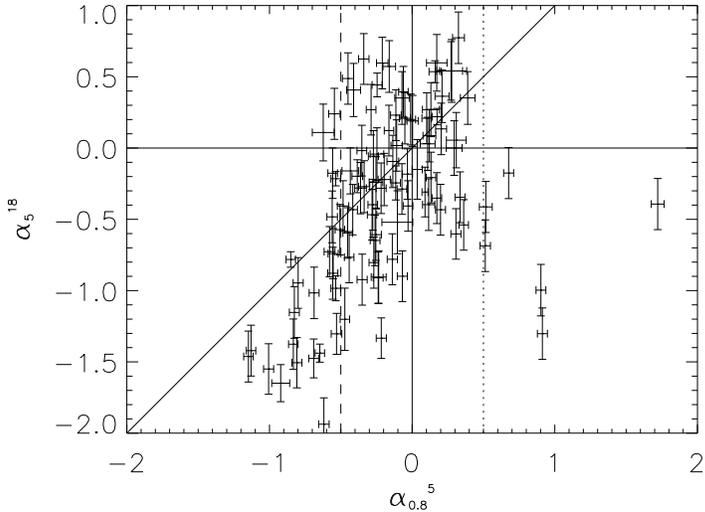} \vspace{7.5cm}
\caption{$\alpha_5^{18}$ vs. $\alpha_{0.8}^5$ scatter plot: the dashed
line represents the separation between flat ($\alpha_{0.8}^5 > -0.5$) and
steep spectrum sources and while the dotted line is the threshold for 
selecting High Frequency Peaker ($\alpha_{0.8}^5 > 0.5$) candidates 
at low frequency.}
\vspace*{-0.5cm}
\label{fig:spixsc}
\end{center}
\end{figure}

Within the boundaries in Fig.~\ref{fig:galcoord} there are 12 WMAP sources.
We found that
six of them fall in small survey ``holes'' where the survey sensitivity
was too poor or array antennas were unable to sample sky positions and 
amplitudes. The effect of these regions with poor or null sensitivity on the
survey effective area and completeness is quantitatively taken in 
account in Fig.~\ref{fig:Aeff}.
For the six sources having a S/N $>$ 4 in the survey we were 
able to produce wide range (0.8--61 GHz) radio spectra which in four cases 
turned out to be steeply rising spectra. One source (PMNJ1803--6507)
could be a HFP candidate possibly peaking at about 40 GHz 
(Fig.~\ref{fig:wmap}). Note that WMAP and 18 GHz pilot survey flux densities 
were measured within a few months of each other (i.e. they are 
near-simultaneous), the PMN (5 GHz) and 
SUMSS (0.843 GHz) measurements date from 10 and 4 years earlier
respectively.         

\begin{figure}
\begin{center}
\includegraphics{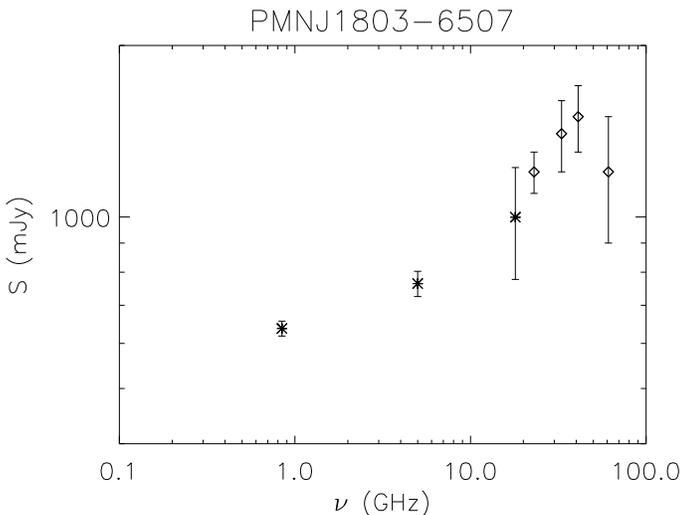} \vspace{7.5cm}
\caption{Radio spectrum spanning between 0.843 and 61 GHz of one of the 
6 sources in the 18 GHz pilot survey present also into the WMAP 
discrete source catalogue. The WMAP points are represented by diamonds.}  
\vspace*{-0.5cm}
\label{fig:wmap}
\end{center}
\end{figure}

\section{Optical identifications of extragalactic sources}

\subsection{Optical identifications from SuperCosmos}
We used the SuperCOSMOS online database (Hambly et al.\ 2001) to
search for optical counterparts to the 18\,GHz sources detected in
the pilot study. Our radio position errors are typically 3.0\,arcsec
in RA and 1.4\,arcsec in Dec (\S7.1).
The optical positions listed in the SuperCOSMOS catalogue also
have an associated uncertainty of up to 1\,arcsec, so we chose a
cutoff radius of 8\,arcsec around the radio position to search for
an optical counterpart. Table 4 lists the results of
this search for the 123 extragalactic sources in our sample.  Of these,
74 (60\%) are identified with stellar objects (candidate QSOs), 30
(25\%) with galaxies and 19 (15\%) are blank fields, i.e. they
have no optical counterpart  in SuperCOSMOS.
Some notes on individual sources are given in Appendix A.

The table columns are as follows: \\
(1) 18\,GHz source name, as in Table 2. \\
(2,3) 18\,GHz radio position (J2000.0) measured from the follow-up images
as described in \S5.1. \\
(4,5) The position (J2000.0) of the closest optical
counterpart identified in the SuperCOSMOS catalogue. Where these columns
are blank, there is no catalogued optical object within 8\,arcsec of the
18\,GHz radio position. \\
(6) Blue (B$_{\rm J}$) apparent magnitude of the optical ID. \\
(7) SuperCOSMOS T parameter, where T=1 corresponds to a
spatially--resolved galaxy and T=2 to a stellar object. \\
(8) Offset (in arcsec) between the radio and optical positions. \\
(9,10) Total (integrated) 18\,GHz flux density and estimated error. 
Note that these are 
generally higher than the peak flux densities listed in Table 2. \\
(11,12) 5\,GHz flux density (in mJy) from the PMN catalogue (Gregory 1994),
and estimated error. \\
(13,14) 843\,MHz flux density (in mJy) from the SUMSS catalogue
(Mauch et al. 2003), and estimated error. Where no flux density is listed,
the source lies in an area of sky where SUMSS observations had not been
completed by mid--2003, and which was therefore not included in the
Jul--1--2003 version of the SUMSS catalogue.\\
(15,16) Radio spectral index between 843\,MHz and 18\,GHz, and estimated
error.\\
(17) Source name in the NASA Extragalactic Database (NED).\\
(18) Notes on NED sources.\\

Sixteen of the objects in Table 4 have published redshifts,
and these range from $z$=0.014 to 0.183 for the galaxies and $z$=0.54
to 3.15 for the QSOs.

Fig.~\ref{fig:id} shows the relation between 18\,GHz flux density
and spectral index for sources identified with QSOs, galaxies and
blank fields.
There appears to be a clear similarity between QSO and galaxy
spectral index distributions, while in low-frequency surveys
these populations have a very different behaviour.
We suggest that our sample is dominated by AGN in galaxy nuclei
rather than radio galaxy lobes. This then points to a close similarity
between radio loud QSO and galaxy nuclei populations.

\subsection{Reliability of the optical IDs}
We used a series of Monte Carlo tests to estimate the number of
spurious SuperCOSMOS radio--source IDs which arise from the chance
superposition of a radio source and a foreground star or galaxy.
To do this, we offset the position of each radio source by an amount
(typically 10--30\,arcmin) which is significantly larger than our
8\,arcsec matching radius, but small enough that the surface density
of SuperCOSMOS stars and galaxies should be similar.  We then
cross--matched the offset radio positions with the SuperCOSMOS
catalogue in the same way as was done for the genuine radio positions.
This was done four times, with offsets to the north, south, east and west
of the original positions, to measure the average number of spurious matches.

We found that for a cutoff radius of 8\,arcsec and an optical magnitude
B$_{\rm J}<22$\,mag, roughly 17\% of the optical IDs in Table 4 are
likely to be spurious, giving an overall reliability of $\sim$83\%.
This could be improved by measuring more accurate radio positions for the
18\,GHz sources, so that the cutoff radius for matching could be reduced.
The small number of very faint optical IDs (B$_{\rm J}=22-23$\,mag)
in Table 4 is close to the number expected by chance, suggesting that
almost all of these very faint objects are chance associations rather than
genuine IDs.

\subsection{Optical identification rate}

Restricting the sample to a limiting magnitude of B$_{\rm J}=22.0$
where the SuperCOSMOS catalogue is complete and the reliability
of the radio--optical matching high, we find an overall optical
ID rate of 80$\pm$8\% (98/122 extragalactic sources).
When corrected statistically for chance coincidences 
with foreground objects, this drops to
73\% with an optical ID brighter than B$_{\rm J}=22.0$, where 
51\% are stellar objects (candidate QSOs), 22\% galaxies and 27\% 
faint objects or blank fields. 
This rate is comparable with the 83\% ID rate reported
by Snellen et al.\ (2002) for a sample of flat--spectrum radio sources
selected at 6\,cm and with flux densities above 200\,mJy.
It is however much higher than that found for weak low-frequency 
samples (Windhorst, Kron \& Koo 1984; McMahon et al. 2002). 




\begin{figure}
\begin{center}
\includegraphics{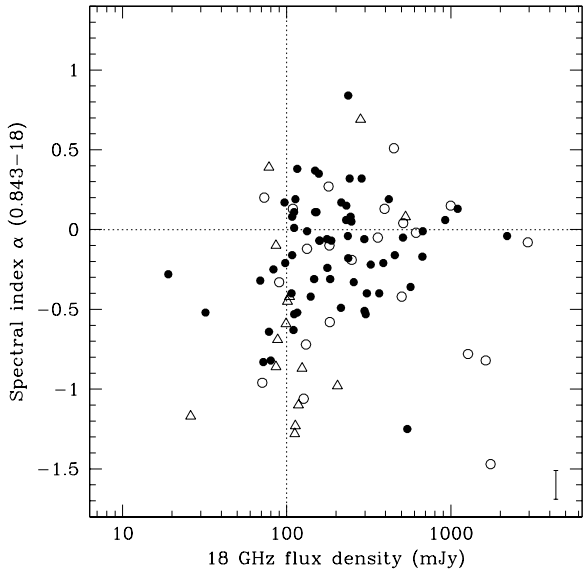}
\vspace{8cm} \caption{Plot of 0.843--18\,GHz spectral index versus
18\,GHz flux density for sources identified as QSOs (filled
circles) and galaxies (open circles), and for sources which are
blank fields ($B_{\rm J}>22$\,mag, triangles). The error bar in the bottom 
right hand corner is representative of the typical uncertainty in the 
spectral indices which is dominated by the typically 20\% errors of 18GHz 
flux densities.} \vspace*{-0.5cm}
\label{fig:id}
\end{center}
\end{figure}

\section{Conclusions}

The wide-band interferometer (total bandwidth spanning from 16 to
20 GHz) produced by linking a wide--band correlator to
two ATCA antennas equipped with low--noise receivers
allowed us to survey the sky at a rate of more than 12\,deg$^2$
per hour to a detection limit of $\sim100$\,mJy despite the small
primary beam area ($\hbox{FWHM}=2.3$\,arcmin). A total of 221
sources have been confirmed by follow-up observations and had flux
densities and positions measured.

The distribution with Galactic latitude shows a
sharp peak with FWHM in $|b|$ of 1.9$^{\circ}$ due to Galactic 
sources.  


We exclude the galactic plane and use only the 123 sources 
with $|b| > 5^\circ$. 
Although this pilot survey is not extremely uniform we know the 
sensitivity map and can estimate the 18 GHz counts of extragalactic 
sources down to
$\simeq 100\,$mJy. The results are in good agreement with the 15
GHz counts by Taylor et al. (2001), with the preliminary VSA
(Grainge et al. 2003) and DASI (Leitch et al. 2002) counts,
with the WMAP counts at 41 GHz (Bennett et al. 2003) and with the
9C survey at 15 GHz by Waldram et al. (2003). We determine the source 
counts particularly well in the crucial 200 - 800mJy flux range,
bridging the gap between the 9C and the WMAP surveys.  
With our estimate of the slope as S$^{-2.2}$ we can confirm  
that the Planck mission will only need to measure and correct for 
foreground sources down to the 100mJy level in order to push the 
fluctuations due to extragalactic radio sources below the instrumental 
noise (cf. Fig.~\ 6 of De Zotti et al. 1999a). 
At this flux level a blind survey of the entire sky at 20GHz 
is feasible.  

Although this survey has not revealed any new class of radio sources, 
we drew attention to the huge scatter in the flux - flux scatter plot 
(Fig.~\ref{fig:ffsc}) which suggests that care is needed  
when using low-frequency surveys to correct high frequency CMB observations.

The radio spectral 
index distribution of the extra-galactic sources shows that over
70\% of them are ``flat''-spectrum ($\alpha_{0.843}^{18}> -0.5$,
$S_{\nu}\propto\nu^{\alpha}$), including 30 (29\%) sources with
rising spectrum ($\alpha_{0.843}^{18} > 0$). We have a significant 
population of sources with minima in their spectrum in the 1-10 GHz range.

The optical identification rate of 73\% is very high: 
51\% of the extra-galactic sample are identified with 
stellar objects (candidate QSOs) and 22\% with
radio galaxies. This is almost a factor of 3 higher 
than the identification rate in surveys at comparable flux levels at 
1.4GHz.  This indicates that the high-frequency source population  
has a luminosity function which does not extend to such high power as 
in lower frequency samples. We thus expect that high-frequency surveys  
will be more sensitive to clustering effects. 

A statistical estimate of fractional polarization distribution yielded 
a median value of 2.1\% which is low compared with the steeper 
spectrum population.  It agrees with the median polarization 
measured for the flat-spectrum population in the K\"uhr sample 
(Ricci et al. 2004) again indicating that this high-frequency survey has not 
revealed any new population of sources at this flux level.  
With the median polarization below that expected for CMB E-mode, 
Planck will have no difficulty correcting for this foreground population.
However B-mode CMB polarization is much harder to detect and 
correction for polarised sources will be necessary. 
Future targeted follow--up with much higher polarization 
sensitivity is possible.

In the near future we plan to cover 10$^4$ deg$^2$
in the southern sky by using an improved and more powerful version
of the wide band correlator (3 baselines and 8 GHz bandwidth)
being able to reach a 100\% completeness for all sources with
S$_{\rm 20 GHz} > 40$ mJy and we expect to detect $\simeq 10^4$
extragalactic sources down to that limiting flux density. 

This survey has confirmed a lower than expected density of sources 
at high frequency making the mm array calibration problem more difficult,
however the full 18GHz ATCA survey will still be able to 
provide a superb calibration source network for ALMA with a density of 
about one source per square degree down to 40mJy.  

\section{Acknowledgments}
We are grateful to the referee for his very careful reading
of the manuscript and his many useful suggestions.
This research was supported in part by the Italian Space Agency
(ASI) and by the Italian MIUR through a COFIN grant. RR gratefully
acknowledges a financial contribution from the Italian National
Research Council (CNR) in the framework of an exchange program
with CSIRO; he also warmly thanks the Paul Wild Observatory staff
for their kind hospitality at Narrabri (NSW, Australia) where part
of this work was carried out. The Australia Telescope is funded
by the Commonwealth of Australia for operation as a National 
Facility managed by CSIRO. This research has made use of the 
NASA IPAC Extragalactic Database (NED) which is operated by the 
Jet Propulsion Laboratory, California Institute of Technology,
under contract with the National Aeronautics and Space Administration.

\newpage

\appendix

\section[]{Notes on individual sources}

{\bf PKS\,0021$-$686:} We adopt the SuperCOSMOS classification as
a stellar object, but note that the 2MASS Extended Objects catalogue
(Jarrett et al. 2002)
classified this as a galaxy with K$_{\rm s}$(total)=13.65\,mag.
Detected as an X-ray source in the RASS Bright Source
Catalogue (Voges et al.\ 1999).

{\bf PKS\,0022$-$60:} Compact double source shown in Fig.\ A1.
The catalogued MRC flux density of 3.83\,($\pm$0.17)\,Jy at 408\,MHz
implies a radio spectral index $\alpha^{0.408}_{18} = -0.78\pm0.02$.

{\bf PKS\,0101$-$649:} The 843\,MHz flux density is from Jones \& McAdam
(1992). Listed in the 2MASS Extended Objects catalogue as a galaxy with
K$_{\rm s}$(total)=13.41\,mag.

{\bf PKS\,0235$-$618:} Detected as an X-ray source in the RASS 
Bright Source Catalogue. Jones \& McAdam (1992) note that the original 
Parkes detection is resolved into a pair of unrelated sources at 843\,MHz.  
The source detected at 18\,GHz is associated with the western (brighter) 
member of the pair.

{\bf PMNJ\,0422$-$6507:} The optical ID is offset 7.9\,arcsec in RA from
the 18\,GHz radio position, but is within 0.5\,arcsec of the SUMSS position.
We accept this as the correct ID because of the close agreement with
the SUMSS position, and because the 18\,GHz source is close to the edge
of the observed field.

{\bf PKS\,0516$-$621:} The optical ID is within 0.1\,arcsec of the
VLBI position measured by Ma et al.\ (1998).

{\bf PKS\,0522$-$611:} Noted as a ROSAT X-ray source by Brinkmann, Siebert
\& Boller (1994).

{\bf PKS\,1105$-$680:} The optical ID found in SuperCOSMOS was first
identified by Jauncey et al. (1989), and is within 0.1\,arcsec of the
Ma et al.\ (1998) VLBI position. The foreground Galactic extinction
in this field ($b=-7.4^\circ$) is A$_{\rm B}=0.99$\,mag.

{\bf PKS\,1133$-$681:} There is no optical object within 5\,arcsec of the
18\,GHz radio position in the SuperCOSMOS catalogue.  White et al.\ (1987)
identify this source with a faint (B=22.2\,mag.) galaxy which is just
visible on the SuperCOSMOS B image.  The foreground Galactic extinction
in this field ($b=-6.6^\circ$) is A$_{\rm B}=2.0$\,mag.


{\bf PKS\,1420$-$679:} The SuperCOSMOS image shows no optical object
within 5\,arcsec of the radio position.  White et al.\ (1987) identify
this source with a faint (B=22.2\,mag.) stellar object, but this should
be regarded as uncertain without spectroscopic confirmation because of
the high surface density of foreground stars at this low Galactic latitude.
The Galactic extinction in this field ($b=-6.8^\circ$) is
A$_{\rm B}=2.4$\,mag.

{\bf PKS\,1448$-$648:} The SuperCOSMOS image shows no optical object
within 5\,arcsec of the radio position.
White et al. (1987) identify this source with a faint (B=22.0\,mag.)
galaxy which is 7.5\,arcsec from the 18\,GHz position, but this object
is classified as stellar by SuperCOSMOS and may be a foreground star.
The extinction in this low-latitude region ($b=-5.1^\circ$) is
A$_{\rm B}=3.2$\,mag.

{\bf WKK\,5585:} Detected as an X-ray source in the RASS Bright Source
Catalogue (Voges et al.\ 1999).

{\bf NGC\,6328:} This nearby ($z$=0.014) galaxy is the optical counterpart
of the radio galaxy PKS\,1718$-$649, which is the closest-known GPS
radio source (Tingay et al. 1997).

{\bf PKS\,1801$-$702:} A bright elliptical galaxy without a catalogued
redshift, K$_{\rm s}$(total)=13.65\,mag in the 2MASS Extended Objects
catalogue.

{\bf PKS\,1814$-$63:} A nearby ($z$=0.0627) active galaxy, with
K$_{\rm s}$(total)=13.65\,mag in the 2MASS Extended Objects catalogue.

{\bf PKS\,1819$-$67:} Highly-polarized double source with no optical
ID, discussed in \S5.3.

{\bf PKS\,1922$-$62:} This source is 4.5\,arcmin from the
strong radio source PKS\,1922$-$62. Large pointing errors in PMN (Parkes 
telescope beam width is about 5 arcmin) are likely to be the cause
of the offset. 

{\bf PKS\,1934$-$638:}  This strong and well--studied radio galaxy 
was the prototype GPS source (O'Dea et al. 1991).

{\bf PMNJ\,2008$-$6110:} Compact source with a highly--inverted radio
spectrum and no optical ID.

{\bf MRC\,2041$-$617:} Detected as an X-ray source in the RASS 
Bright Source Catalogue (Voges et al. 1999) and shown in Fig.\ A1.

{\bf ESO\,075$-$G41:}  The two components detected here (Fig.\ A1)
are the core and southern hotspot of this nearby ($z$=0.028) radio galaxy.

{\bf PKS\,2210$-$637:} Detected as an X-ray source in the RASS 
Bright Source Catalogue (Voges et al. 1999).

{\bf PKS\,2300$-$683:} Noted as a ROSAT X-ray source by Brinkmann,
Siebert \& Boller (1994).

{\bf PKS\,2303$-$656:} There is no optical counterpart in the SuperCOSMOS
catalogue, but Jackson et al.\ (2002) report that a faint red galaxy
is visible on CCD images of this field, with a tentative redshift of
$z$=0.470 from an optical spectrum.

{\bf IRAS\,23074$-$5957:} Noted by Roy \& Norris (1997) as a member of
the rare class of `radio--excess infrared galaxies'.
Listed in the 2MASS Extended Objects catalogue with
K$_{\rm s}$(total)=14.00\,mag.

{\bf PKS\,2356$-$61:} Well known FRII radio galaxy (Koekemoer \& 
Bicknell 1998)-- both hotspots are detected at 18\,GHz.  
The host galaxy has K$_{\rm s}$(total)=12.86\,mag 
in the 2MASS Extended Objects catalogue.

\begin{figure*}
\centering
\begin{minipage}{180mm}

\vspace*{6.5cm}



{\large \hspace*{1.5cm}(a) PKS\,0022$-$60} \hspace*{4.0cm}
{\large (b) PMNJ\,2008$-$6110}

\vspace*{6.5cm}

{\large \hspace*{1.5cm}(c) MRC\,2041$-$617} \hspace*{4.0cm}
{\large (d) ESO\,075$-$41}

\vspace*{6.5cm}


{\large \hspace*{0.1cm}(e) PKS\,2356$-$61} \hspace*{4.0cm}
{\large (f) PKS\,2356$-$61 with SUMSS 843\,MHz contours overlaid}

\vspace*{0.5cm}
\caption{
Radio contours overlaid on SuperCOSMOS optical images of sources detected at 18\,GHz and discussed in the text.
}
\end{minipage}
\end{figure*}


\begin{figure*}
\centering
\begin{minipage}{180mm}

\vspace*{6.5cm}



{\large \hspace*{1.5cm}(a) PKS\,0021$-$686} \hspace*{4.0cm}
{\large (b) PKS\,0101$-$649}

\vspace*{6.5cm}

{\large \hspace*{1.5cm}(c) PKS\,0235$-$618} \hspace*{4.0cm}
{\large (d) PMNJ\,0422$-$6507}

\vspace*{6.5cm}


{\large \hspace*{0.1cm}(e) PKS\,0516$-$621} \hspace*{4.0cm}
{\large (f) PKS\,0522$-$611}

\vspace*{0.5cm}
\caption{
Radio contours overlaid on SuperCOSMOS optical images of sources detected at 18\,GHz and discussed in the text.
}
\end{minipage}
\end{figure*}


\begin{figure*}
\centering
\begin{minipage}{180mm}

\vspace*{6.5cm}



{\large \hspace*{1.5cm}(a) PKS\,1105$-$680} \hspace*{4.0cm}
{\large (b) PKS\,1133$-$681}

\vspace*{6.5cm}

{\large \hspace*{1.5cm}(c) PKS\,1420$-$679} \hspace*{4.0cm}
{\large (d) PKS\,1448$-$648}

\vspace*{6.5cm}


{\large \hspace*{0.1cm}(e) WKK\,5585} \hspace*{4.0cm}
{\large (f) NGC\,6328}

\vspace*{0.5cm}
\caption{
Radio contours overlaid on SuperCOSMOS optical images of sources detected at 18\,GHz and discussed in the text.
}
\end{minipage}
\end{figure*}


\begin{figure*}
\centering
\begin{minipage}{180mm}

\vspace*{6.5cm}



{\large \hspace*{1.5cm}(a) PKS\,1801$-$702} \hspace*{4.0cm}
{\large (b) PKS\,1814$-$63}

\vspace*{6.5cm}

{\large \hspace*{1.5cm}(c) PKS\,1819$-$67} \hspace*{4.0cm}
{\large (d) SUMSSJ\,192659$-$624225}

\vspace*{6.5cm}


{\large \hspace*{0.1cm}(e) PKS\,1934$-$63} \hspace*{4.0cm}
{\large (f) PKS\,2210$-$637}

\vspace*{0.5cm}
\caption{
Radio contours overlaid on SuperCOSMOS optical images of sources detected at 18\,GHz and discussed in the text.
}
\end{minipage}
\end{figure*}


\begin{figure*}
\centering
\begin{minipage}{180mm}

\vspace*{6.5cm}



{\large \hspace*{1.5cm}(a) PKS\,2300$-$683} \hspace*{4.0cm}
{\large (b) PKS\,2303$-$656}

\vspace*{6.5cm}

{\large \hspace*{1.5cm}(c) IRAS\,23074$-$5957} 




\vspace*{0.5cm}
\caption{
Radio contours overlaid on SuperCOSMOS optical images of sources detected at 18\,GHz and discussed in the text.
}
\end{minipage}
\end{figure*}

\end{document}